\documentclass[12pt]{article}
\usepackage{epsf}

\newcount\xrefpos \xrefpos=0
\newcount\yrefpos \yrefpos=0
\newcount\xput \xput=0
\newcount\yput \yput=0
\def\refpos#1 #2 #3{\global\xrefpos=#1 \global\yrefpos=#2
                         \rlap{$\smash{#3}$}}
\def\put #1 #2 #3{\xput=#1 \yput=#2
                  \advance\xput by -\xrefpos
                  \advance\yput by -\yrefpos
                  \rlap{\kern\the\xput truebp
                        \vbox to 0pt{\vss\hbox{$\displaystyle #3$}%
                        \kern\the\yput truebp}}}
\def\beginlabels\refpos#1\endlabels{\hbox{$\refpos#1$}}

\begin{document}

\bibliographystyle{unsrt}

\vspace*{-.6in}
\thispagestyle{empty}
\begin{flushright}
DAMTP R-98/02\\
hep-th/9801127
\end{flushright}
\baselineskip = 20pt

\vspace{.5in}
{\Large\bfseries
\begin{center}
Bound States of D-Branes and the Non--Abelian Born--Infeld Action
\end{center}}
\vspace{.4in}

\begin{center}
D. Brecher\footnote{email d.r.brecher@damtp.cam.ac.uk.} and
M. J. Perry\footnote{email malcolm@damtp.cam.ac.uk.}\\
\emph{D.A.M.T.P., University of Cambridge, Cambridge CB3 9EW, U.K.}
\end{center}
\vspace{1in}

\begin{center}
\textbf{Abstract}
\end{center}
\begin{quotation}
We attempt to settle the issue as to what is the correct
non--abelian generalisation of the Born--Infeld action, via a consideration of the
two--loop $\beta$--function for the non--abelian background
gauge field in open string theory.  An analysis of the bosonic theory
alone shows the recent proposal of Tseytlin's~\cite{tseytlin} to be
somewhat lacking.  For the superstring, however, this proposal would seem to be correct, and not
just within the approximation used in~\cite{tseytlin}.  Since it is
this latter case that is relevant to the description of D-branes we,
in effect, obtain an independent verification of Tseytlin's result.
Some issues involved in the concept of non--abelian T--duality are
discussed; and it is shown how the interaction between separated and
parallel branes, in the form of massive string states, emerges. 
\end{quotation}
\vfil

\newpage

\pagenumbering{arabic}

\section{Introduction}

The connections between D-branes~\cite{dai+leigh+pol,pol:dbrane,pol:tasi} and
the non--linear electrodynamics of Born and Infeld~\cite{born+infeld} are
well--known.  The pure Born--Infeld (BI) action was found to describe
the low energy effective field theory of open bosonic string theory
in~\cite{callan+etal}, using the background field technique applied to
the two--dimensional $\sigma$--model.  These
methods were extended to an analysis of a Dirichlet
$\sigma$--model in~\cite{leigh}, where the Dirac--Born--Infeld (DBI) action
was found to describe the
world--volume theory of the D$p$-brane; this latter can be
deduced easily by a consideration of T--duality as applied to the
BI theory, as we will review below.  Such a theory is one of a
non--linear electrodynamics on the brane, the transverse coordinates of
which are T--dual to the world--volume (abelian) gauge field.  It is
natural to ask what is the non--abelian
generalisation of this theory, and there has been some
recent~\cite{tseytlin,dorn:dbranes}, and some not so
recent~\cite{argyres+nappi,dorn+otto}, work addressing this question.  The issue has not been
resolved fully, however; and it is at just such a resolution that this paper is
aimed.

By attaching charges, the Chan--Paton factors, to the ends of open strings, we
generate a spacetime gauge theory.  By restricting the strings to end
on a $(p+1)$--dimensional hypersurface, the D$p$-brane, we have a gauge
theory on the world--volume of the brane.  Since open strings come in
two different types,
oriented and unoriented, there are different possible gauge
theories: a $U(N)$ theory for the oriented string; and an $SO(N)$ or
$USp(N)$ theory for the unoriented string.  Within D-brane theory, the
former is of interest because a $U(N)$ gauge theory is thought to
describe a
collection of $N$ D$p$-branes~\cite{witten}.  This is observed easily:
if we have $N$ separated and parallel D$p$-branes, there will
be massless open strings ending on each brane, giving a $U(1)^N$ gauge
theory.  There will also be open strings stretched between the branes,
however, with a mass proportional to their length.  As the branes are
brought together, these strings become massless, extending the
resulting gauge group to $U(N)$.  Since, roughly speaking, $U(N) \sim U(1) \times
SU(N)$, such a theory will be one of a bound state of such D-branes, the $U(1)$ factor
corresponding to the collective coordinate; and the action describing such
states ought, then, to be a non--abelian generalisation of the
DBI action.  Interestingly, if we regard this bound state as a `brane'
in its own right, we see that its world--volume theory will contain a
non--abelian gauge field; and the transverse coordinates of the
`brane' will be T--dual to this.  That is, they will themselves become matrix--valued, a notion which has
lead to the Matrix description of M--Theory~\cite{banks+etal}.   By
deciding upon what we believe to be the correct non--abelian BI
action, we hope to be able to answer some of the questions which arise in this context.

Since we will employ the background field technique we give,
in section two, a very brief description of how this method has been
used in open bosonic string theory to derive the results pertaining to
D-branes.  In section three we consider a Yang--Mills background
field, and calculate the $\beta$--function for this field to two--loops,
following closely the analysis of~\cite{dorn+otto}.  Since D-branes
are embedded in superstring theory, as opposed to the purely bosonic
theory, we then extend these results to a consideration of the open
superstring (and thereby provide an explanation for why the
aforementioned purely bosonic results carry through to the full--blown
supersymmetric theory).  In section four we consider the possible non--abelian generalisations
of the BI action, finding one which gives an equation of motion
consistent with the results of section three only.  Finally, we make
some comments as to the nature of non--abelian T--duality; the first
steps towards explaining the appearance of matrix--valued spacetime
coordinates.  For separated and parallel branes, we should see an
interaction between them in the form of massive strings.  To lowest
order this will emerge as a massive Yang--Mills theory; and we show
how this comes about.  We offer further a few remarks concerning D4-branes
with an (anti--)self--dual field strength; one of the few obvious cases for
which the non--abelian BI action is tractable.

\section{Background Field Method in Open String Theory}

It is well--known that consistency\footnote{There
are two ways of looking at this: on the one hand, \emph{conformal}
invariance of the string theory is the requirement that the trace of the energy--momentum
tensor vanish i.e., that the Weyl anomaly coefficients are zero.  This
is how the results of, e.g.~\cite{callan+etal2}, were found.  On the other, \emph{scale}
invariance is the requirement that the renormalisation group
$\beta$--functions vanish.  We are dealing here with the latter
approach although this will not effect the generality of our results,
since there is a well--defined procedure for generating the anomaly
coefficients from the $\beta$--functions, and vice
versa~\cite{callan+thorlacius}.} of tree--level
string theory determines the equations of motion of the background
spacetime in which the string lives, and of the matter fields therein; for an extensive review,
see~\cite{callan+thorlacius}.  The $\sigma$--model is a theory of
two--dimensional massless scalar fields, the spacetime coordinates,
living on
the string worldsheet; the `coupling constants' of the model being the
metric $G_{ab}$, the two--form $B_{ab}$, the dilaton $\Phi$ and, for open strings, the
gauge field $A_a$.  Two--dimensional loop effects contain logarithmic
divergences, so these couplings must be renormalised.  The requirement
that this renormalisation be scale--independent is just the
requirement that the $\beta$--function for each field vanish.  From a stringy
perspective, however, the couplings are just the
classical spacetime fields in which the string lives; condensates of
the massless string states.  The
statement $\beta^{(A)} = 0$ is, then, interpreted as an equation of motion
for the spacetime field $A_a(X)$.

The background field method consists of splitting the embeddings $X^a$ into a classical and a quantum piece,
\begin{equation}
X^a (\tau ) = \overline{X}^a (\tau ) + \xi^a (\tau ),
\label{split}
\end{equation}

\noindent thereby constructing a two--dimensional quantum field theory in the
variable $\xi^a(\tau )$.  $\tau$ is the time coordinate on the
worldsheet, only the $\tau$ dependence being relevant here, since
all gauge field interactions occur on the boundary of the worldsheet.
We work in flat spacetime, thereby avoiding the
geometrical considerations concerning the identification of
the quantum field as a normal coordinate~\cite{alv-gau+freedman}; and, since we are
interested in divergences alone, we work with a euclidean
target space and worldsheet.  Expanding the usual string action
$S[X] = S[\overline{X} + \xi]$ about an arbitrary on--shell reference configuration
$\overline{X}^a$, the
functional generating all loop diagrams with the external legs amputated is
\[
\Omega[X] = \int D\xi \exp \left( - \left(S[\overline{X}+\xi] -
S[\overline{X}] \right)\right),
\]

\noindent with $\hbar=1$.  Following~\cite{alv-gau+freedman} we work
using time--dependent perturbation theory, writing
\[
\Omega[X] = \langle 0| \exp \left( -\left( S[\overline{X} + \xi]
-S[\overline{X} \right)\right) |0
\rangle.
\]

\noindent Then the effective action is
\begin{equation}
\Gamma[X] = -\ln \Omega [X] = \langle 0| \left(S[\overline{X} + \xi] -
S[\overline{X}]\right)|0
\rangle.
\label{effact}
\end{equation}

\noindent All possible loop diagrams are contained in (\ref{effact}), some of which
will be divergent.  To cancel these divergences, counterterms must be
added to the action: $S[X] \rightarrow S[X] + \bigtriangleup S$; and it is via the addition of such counterterms that the spacetime fields get renormalised.

These techniques were used in~\cite{callan+etal} to deal with the open
bosonic string in a background Maxwell field.  It was found that
the solutions of the
equation of motion following from the
BI lagrangian,
\begin{equation}
{\cal L}_{BI} = \sqrt{ \det (\delta_{ab} + 2\pi\alpha' F_{ab})},
\label{lbiold}
\end{equation}

\noindent are just the solutions of the condition for scale invariance,
$\beta^A_a = \alpha'(1 - (2\pi\alpha'F)^2)^{-1bc}\partial_{(b}F_{c)a} = 0$:
\begin{equation}
\frac{\delta {\cal L}_{BI}}{\delta A_a} = \sqrt{ \det (\delta_{ab} +
2\pi\alpha' F_{ab}})~ (1 - (2\pi\alpha'F)^2)_a^{-1b} ~\beta^A_b = 0.
\label{deltalbi}
\end{equation}

\noindent Indeed, as we will comment on below, this is true for the superstring
also~\cite{tseytlin:bi}.  This result is valid to all orders in
$\alpha'$, the two--dimensional loop counting parameter, although is
exact for a slowly--varying field strength only; in the generic case, there will
presumably be corrections depending on derivatives of the field strength.
In the non--abelian case, it does not appear to be possible to obtain an analogously
exact (in $\alpha'$) result, and so we work to the two--loop order
only.  Moreover, and as we will discuss further below, it is unclear
as to how the variation, analogous to (\ref{deltalbi}), carries
through to the non--abelian case.  At the order to which we are
working, however, the analysis would seem to be valid.  

We note here that we will in fact take the BI lagrangian to be
slightly different to (\ref{lbiold}): for a euclidean target space we have
\begin{equation}
{\cal L}_{BI} = (2\pi \alpha'g)^{-2}\left[1 -\sqrt{ \det (\delta_{ab} +
2\pi\alpha'g F_{ab})}\right],
\label{lbi}
\end{equation}

\noindent where $2\pi\alpha'$ is the inverse string tension, and we reintroduced the gauge coupling $g$.  This differs from (\ref{lbiold}) only in that it
is zero for a vanishing field strength.

The pure BI Lagrangian (\ref{lbi}) describes a low energy
effective spacetime theory in which the effects of gravity are ignored: the
open strings, having electromagnetic charges at their ends, generate a
ten--dimensional spacetime gauge theory.  To make contact with the concept of
the D-brane, we must consider T--dualising this theory~\cite{pol:tasi}.
That is, if we compactify the theory on a circle of radius $R$ in, say, the $X^9$
direction, we can
reformulate it in terms of the dual coordinate, $\tilde{X}^9$. Now,
however, we find the ends of the strings to lie on the same plane,
$\tilde{X}^9(\pi)-\tilde{X}^9(0) = 2\pi n \alpha'/R =
2\pi n\tilde{R}$: a D-brane has appeared on the dual circle, of radius
$\tilde{R}=\alpha'/R$.  Moreover, we have that
$\tilde{X}^9=2\pi\alpha'g A_9$: the gauge field in the compact
direction is T--dual to the transverse coordinate of the D-brane.  In
the dual picture, then, we have $F_{a9}=(2\pi\alpha'g)^{-1}\partial_a
\tilde{X}_9, ~~a, b\neq 9$.  Using the identity~\cite{gibbons}
\begin{equation}
\det\left( \begin{array}{cc} N & -A^t \\ A & M
\end{array}\right)=\det(M)\det(N+A^tM^{-1}A)=\det(N)\det(M+AN^{-1}A^t),
\label{identity}
\end{equation}

\noindent with $t$ denoting transposition, we have the well--known
result
\begin{eqnarray}
{\cal L}_{DBI}= (2\pi\alpha'g)^{-2}\left[ 1 - \sqrt{\det (\delta_{ab} +
\partial_a X^9 \partial_b X^9 + 2\pi\alpha'g F_{ab} )} \right],
\end{eqnarray}

\noindent where we have dropped the tilde on the dual coordinate.
This is the DBI Lagrangian in the so--called static gauge, describing
a D8-brane, $X^9$ being its transverse coordinate;
$\delta_{ab}+\partial_aX^9\partial_bX^9$ is just the pull--back to the
worldvolume of the spacetime metric.  It is important to
realise that this is the T--dual of the pure BI spacetime theory
(\ref{lbi}).  In the same way, the ten--dimensional non--abelian generalisation of
the BI theory will describe an effective spacetime non--abelian gauge theory; and
it is only through T--dualising this that we will find bound states of D-branes.

\section{$\beta$--Functions for Pure Yang--Mills Backgrounds}

Since we are not concerned with effects in the bulk of the worldsheet,
we set $\Phi = B_{ab} = 0$.  The action for open superstrings coupled
to a Yang--Mills spacetime field $A_a(X)$ is then
\begin{eqnarray}
S &=& 
S_{\Sigma} + S_{\partial\Sigma}, \nonumber\\
S_{\Sigma} 
  &=& 
\frac{1}{4\pi\alpha'} \int_{\Sigma} d^2\sigma \left[ \sqrt{\gamma}
\gamma^{\mu\nu} \partial_{\mu}X^a \partial_{\nu}X_a - \frac{i}{2}
\overline{\Psi}^a \rho^{\mu} \partial_{\mu} \Psi_a \right], \label{sint}\\
S_{\partial\Sigma}
  &=&
-\ln {\rm Tr} ~  {\cal P} \exp \left( ig
\oint_{\partial\Sigma} d\tau \left[ A_a (X)\partial_{\tau}X^a
-\frac{1}{2} \psi^a \psi^b F_{ab}(X) \right] \right) \label{sbound}\\
  &=&
-\ln {\rm Tr} ~ {\cal P} \left(U[A] \right), \nonumber
\end{eqnarray}

\noindent where $-\infty < \tau < \infty$, $0 \leq \sigma
\leq \pi$, $\mu$ is a worldsheet
index, $\rho^{\mu}$ being the two--dimensional Dirac matrices, and
$a,b=0,\ldots,9$.  The spinor $\psi^a(\sigma, \tau)$ in (\ref{sbound})
is just the restriction to the boundary of the usual worldsheet
Majorana spinor $\Psi^a(\sigma, \tau)$: $\psi^a =
\Psi^a\left.\right|_{\partial\Sigma}$.  That is, it depends only on
the combination of the left and right moving fermions which is not set to zero
by the boundary conditions.  The gauge field $A_a(X)=A_a^i(X)t^i$
takes values in the fundamental representation of the group algebra, which we leave unspecified for the
time being, and $F_{ab}(X) = \partial_a A_b - \partial_b A_a -ig[A_a,
A_b]$ is the non--abelian field strength.  The factor of $i$ in (\ref{sbound}) is necessary if
$\{t^i\}$ is to be an hermitian basis of the group algebra.  ${\cal P}$ is
the path--ordering operation and ${\rm Tr}~{\cal P} \left( U[A]
\right)$ is the supersymmetrised Wilson loop.  The path--ordered exponential reduces to a standard
exponential if the matrices at different spacetime positions commute i.e., if
the gauge group is abelian; in which case we would have the usual
$S_{\partial\Sigma} =  -ig \oint_{\partial\Sigma} d\tau \left[ A_a(X)
\partial_{\tau}X^a -\frac{1}{2} \psi^a \psi^b F_{ab}(X) \right]$.

We must note that the background--quantum split (\ref{split}) is to be
applied to the (bosonic) embeddings only, the fermionic variable
$\psi^a$ being treated as a quantum field from the outset.  Making
this split, then, and
expanding (\ref{sbound}) with respect to $\xi^a$ we have
\[
\exp\left(-(S_{\partial \Sigma}[\overline{X} + \xi] - S_{\partial
\Sigma}[\overline{X}])\right) = {\rm Tr}~{\cal P} \left(
U[A(\overline{X}+\xi)] - U[A(\overline{X})] \right) =
\]
\[
ig \oint_{\partial \Sigma} d\tau~ {\rm Tr} ~ {\cal P} \left(
U[A] \left[ F_{ab} \xi^a \partial_{\tau} \overline{X}^b
-\frac{1}{2}F_{ab}\psi^a\psi^b -\frac{1}{2}
D_a F_{bc} \xi^a \psi^b \psi^c + \frac{1}{2}
D_b F_{ac} \xi^a \xi^b \partial_{\tau} \overline{X}^c
\right. \right.
\]
\[
+ \frac{1}{2}
F_{ab} \xi^a \partial_{\tau} \xi^b-\frac{1}{4} D_a D_b F_{cd} \xi^a
\xi^b \psi^c \psi^d + \sum_{n=3}^{\infty} \left( \frac{1}{n!} D_{a_1} \ldots D_{a_{n-1}}
F_{{a_n}b} \xi^{a_1} \ldots \xi^{a_n} \partial_{\tau} \overline{X}^b
\right.
\]
\begin{equation}
+ \left. \left. \left. \frac{n-1}{n!} D_{a_1} \ldots D_{a_{n-2}}
F_{a_{n-1}a_n} \xi^{a_1} \ldots \xi^{a_{n-1}} \partial_{\tau} \xi^{a_n}
-\frac{1}{2}\frac{1}{n!} D_{a_1} \ldots D_{a_n} F_{bc} \xi^{a_1}
\ldots \xi^{a_n} \psi^b \psi^c \right) \right] \right),
\label{expansion}
\end{equation}

\noindent where $D_a$ is the gauge covariant derivative, and all spacetime
fields are evaluated at $\overline{X}^a$.  We have kept the $F_{ab}\psi^a
\psi^b$ term for the reasons mentioned above.  We include the derivation
of (\ref{expansion}) in an Appendix for completeness (see,
e.g.,~\cite{gervais+neveu,dorn:wilsonloops}).  We will ignore the
${\cal O} (\xi)$ terms which vanish on--shell, thereby simplifying our
analysis as compared to that of~\cite{dorn+otto}.  All interactions
relevant to $\beta^{(A)}$ are contained in the expansion
(\ref{expansion}), via the expression (\ref{effact}) for the effective
action; those interactions occurring in the interior of the worldsheet,
generated by (\ref{sint}), effect a
renormalisation of the spacetime metric only.

\subsection{The Bosonic Sector}

Setting the fermions to zero, we have $\exp \left( -S_{\partial
\Sigma}[X] \right) = {\rm Tr}~{\cal P} (U[A])$ being the standard Wilson
loop.  We cannot define the exact propagator of the theory, due to the
path--ordering involved in this boundary term which is why, in
contrast to the abelian case, we cannot work to all orders in
$\alpha'$, even for a slowly--varying field strength.  Expanding the interior
action (\ref{sint}) with respect to $\xi^a$ and requiring the ${\cal
O}(\xi)$ term to vanish gives the usual equation of motion $\partial_{\mu} (\sqrt{\gamma}
\gamma^{\mu\nu}\partial_{\nu}) \overline{X}^a = 0$ and Neumann
boundary condition $ \left.\partial_n \overline{X}^a \right|_{\partial
\Sigma} = 0$.  Due to conformal invariance of the classical theory, we
are free to work on the unit disc, with the coordinates
$z=\tau + i\sigma$ and $\overline{z}=\tau - i\sigma$.  Then the
propagator $G^{ab}(z,z') = \langle 0 |T\left[ \xi^a(z) \xi^b(z')
\right]| 0 \rangle = -2\pi\alpha'\delta^{ab} N(z,z')$ where $N(z,z')$
is the Neumann function on the disc,
\noindent
\begin{equation}
N(z,z') = \frac{1}{2\pi}\ln \left( |z-z'||z- \overline{z}'^{-1} | \right),
\end{equation}

\noindent satisfying $\partial_{\mu} (\sqrt{\gamma}
\gamma^{\mu\nu}\partial_{\nu}) N(z,z') = \delta^2(z-z')$.

On the boundary of the worldsheet we can work with the single angular
variable $\theta, ~~0\leq\theta\leq 2\pi$, and we have~\cite{frad+tseytlin}
\begin{eqnarray}
N(z,z') = N(e^{i\theta}, e^{i\theta'}) &=& \frac{1}{\pi} \ln |z - z'|
 \nonumber \\
 &=&
\frac{1}{2\pi} \ln (2-2\cos\beta), \nonumber \\
 &=&
-\frac{1}{\pi} \sum_{n=1}^{\infty} \frac{\cos n\beta}{n} e^{-\varepsilon
 n},
\label{Nbound}
\end{eqnarray}

\noindent where $\beta = \theta -\theta'$ and $\varepsilon$ is an
ultra--violet cut--off~\cite{tseytlin}.  Note that we
need not concern ourselves over infra--red divergences: since they can
be
regulated via $G \sim \ln(\mu|z - z'|)$, $\mu$ an infra--red cut--off, they will
not contribute in the limit $z \rightarrow z'$.

Counterterms must be added to (\ref{sbound}), then, to cancel the ultra--violet
divergences arising from the propagator
$G^{ab} (0) = -2\pi\alpha'\delta^{ab} N(0) = -2\pi\alpha'\delta^{ab} \left(
N_{div} + N_{fin} \right)$ with, from (\ref{Nbound}), $N_{div} =
\frac{1}{\pi}\ln (\varepsilon) \left. \right|_{\varepsilon \rightarrow
0}$.  From (\ref{sbound}), these counterterms must take the form
\[
\exp(-\bigtriangleup S) = {\rm Tr} ~{\cal P}(U[A + \delta A] - U[A]).
\]

\noindent Since $G^{ab} \sim \alpha'$, $\alpha'$ is
the loop counting parameter for the two--dimensional field theory, and
we write
\[
\delta A_a = \alpha' \delta_1 A_a + \alpha'^2 \delta_2 A_a + {\cal
O}(\alpha'^3).
\]

\noindent Then the counterterms are generated via
\[
\exp(-\bigtriangleup S) = {\rm Tr} ~{\cal P} (U[A +
\alpha'\delta_1 A + {\alpha'}^2\delta_2 A] - U[A]) =
\]
\[
i\alpha' \int_0^{2\pi} d\theta ~{\rm Tr}~{\cal P} \left( U[A]
~\delta_1 A_a \partial_{\theta} \overline{X}^a \right) + i\alpha'^2
\int_0^{2\pi} d\theta ~{\rm Tr}~{\cal P} \left( U[A] ~\delta_2 A_a
\partial_{\theta} \overline{X}^a \right)
\]
\begin{equation}
-\frac{1}{2}{\alpha'}^2 \int_0^{2\pi} d\theta d\theta' ~{\rm Tr}~{\cal P} \left(
U[A] ~\delta_1 A_a(\overline{X}(\theta)) \partial_{\theta}
\overline{X}^a ~ \delta_1 A_b (\overline{X}(\theta'))
\partial_{\theta'} \overline{X}^b \right) + {\cal O}(\alpha'^3).
\label{counterexp}
\end{equation}

\noindent Scale invariance of the theory is guaranteed by requiring the
$\beta$--function for the Yang--Mills field to vanish:
\[
\beta_a = \beta_a^i t^i = \frac{\partial}{\partial (\ln \varepsilon)} \delta A_a =
0,
\]

\noindent this being interpreted as the equation of motion for $A_a$.

Contributions at the 1--loop level come from the ${\cal O}
(\xi^2)$ terms in the expansion (\ref{expansion}), giving the diagrams
of Fig. 1.  Fig. 1(b) is identically zero, since the derivative of the
propagator with respect to the boundary variable $\theta$ is
$\partial_{\theta} G (0) = 0$.  This leaves Fig. 1(a)
alone, which generates
\begin{equation}
\exp(-\bigtriangleup S_1) =  ig \pi \alpha' N_{div} \int_0^{2\pi} d\theta~{\rm
Tr} ~  {\cal P} \left( U[A]~D^b F_{ba}  \partial_{\theta} \overline{X}^a \right).
\label{counterterm1}
\end{equation}
\newline
\[
\vbox{
         \beginlabels\refpos 76 256 {}
                     \put 168 236 {(a)}
                     \put 340 236 {(b)}
                     \put 205 132 {DF\partial_{\tau} \overline{X}}
                     \put 392 132 {F}
         \endlabels
         \epsfxsize=.5\hsize
         \centerline{\epsfbox{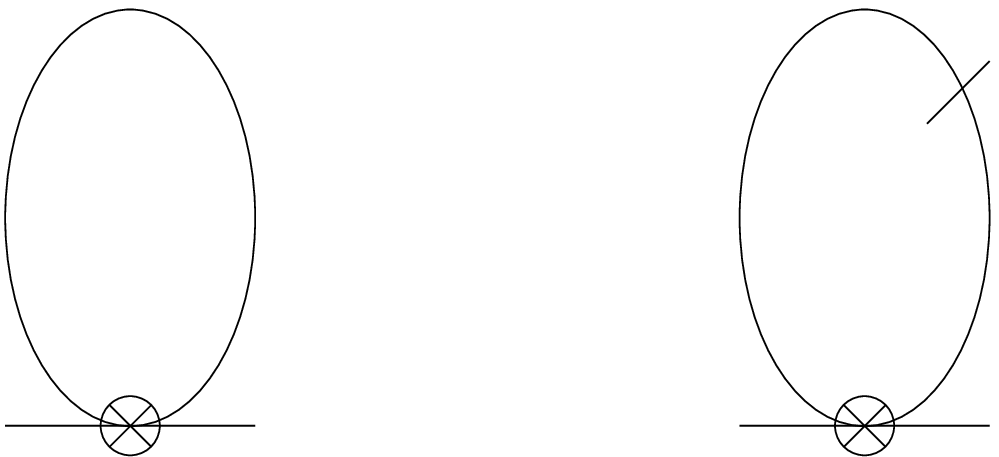}}
         \vspace{15pt}
         \centerline{\bf \footnotesize Figure 1.  One--loop diagrams.}
         \vspace{-.1in}
         \centerline{\footnotesize The solid loops represent a bosonic propagator;
         and a slash, the derivative of the propagator.}
         \vspace{-.1in}
         \centerline{\footnotesize Due to the symmetry properties
         invloved, (b) is identically zero.}
            }
\]
\newline
\noindent Comparison with (\ref{counterexp}) implies
\begin{equation}
\delta A_a^i = \alpha'g D^b F_{ba}^i \ln (\varepsilon) + {\cal O}({\alpha'}^2),
\label{deltaA1}
\end{equation}

\noindent and so
\begin{equation}
\beta_a^i = \alpha'g D^b F_{ba}^i + {\cal O}({\alpha'}^2).
\label{beta1}
\end{equation}

\noindent To lowest order, then, $\beta_a^i = 0$ is just the familiar
Yang--Mills field equation.

At the two--loop level, there are two distinct contributions: those terms
through ${\cal O} (\xi^4)$ in (\ref{expansion}); but also those through
${\cal O} (\xi^2)$ in an iterative expansion of the first order
counterterm (\ref{counterterm1}).  Again ignoring ${\cal O} (\xi)$
terms, the latter is given by
\[
\exp\left(-\bigtriangleup S_1[\overline X + \xi]\right) = ig\pi\alpha'\frac{1}{2}N_{div}
\int_0^{2\pi} d\theta ~{\rm Tr}~{\cal P} \left( U[A] ~\left[
(D_a D^c F_{cb} - D_b D^c F_{ca}) \xi^a \partial_{\theta} \xi^b
\right.\right.
\]
\begin{equation}
+\left.\left.(D_b D_d D^c F_{ca} - D_a
D_b D^c F_{cd}) \xi^b \xi^d \partial_{\theta} \overline X^a \right] \right),
\label{expansion2}
\end{equation}

\noindent which generates the on--shell diagrams of Fig. 2.  Fig. 2(b)
being identically zero, we are left with 
\[ 
\exp(-{\rm Fig. ~2(a)}) = 
\]
\begin{equation}
-ig(\pi\alpha')^2 N_{div} \int_0^{2\pi} d\theta ~ N(0)~{\rm
Tr} ~ {\cal P} \left( U[A]~\left[ D^2 D^b F_{ba} - D^c D_a D^b F_{bc}
+ ig [D^b F_{bc}, F^c_{~a}] \right] \partial_{\theta} \overline{X}^a
  \right),
\label{fig2}
\end{equation}

\noindent where the relation $[D_a, D_b] F_{cd} = -ig[F_{ab},
F_{cd}]$ has been used.  
\newline
\newline
\[
\vbox{
         \beginlabels\refpos 75 435 {}
                     \put 170 420 {(a)}
                     \put 345 420 {(b)}
                     \put 170 310 {(DDF-DDF)\partial_{\tau} \overline{X}}
                     \put 320 310 {(DDDF-DDDF+[D,D]DF)}
         \endlabels
         \epsfxsize=.5\hsize
         \centerline{\epsfbox{myfig1.eps}}
         \vspace{15pt}
         \centerline{\bf \footnotesize Figure 2.  Two--loop diagrams.}
         \vspace{-.1in}
         \centerline{\footnotesize  Diagrams generated from an
         iterative expansion of the first order counterterm.}  
         \vspace{-.1in}
         \centerline{\footnotesize Due to the 
         symmetry properties involved, (b) is identically zero.}
            }
\]

The two--loop diagrams generated by
(\ref{expansion}) are shown in Fig. 3.  Note that the ${\cal O} (\xi^3)$ terms do not
contribute at all, since we have consistently dropped the ${\cal
O}(\xi)$ terms.  Then
\[
\exp(-{\rm Fig. ~3(a)}) = 
\]
\begin{equation}
ig \frac{1}{6}(\pi\alpha')^2 \int_0^{2\pi} d\theta~[N(0)]^2~ {\rm Tr} ~ {\cal P} 
\left( U[A]~ \left[D^2 D^b F_{ba} + D^c D^b D_c F_{ba} + D^b D^2 F_{ba}
\right] \partial_{\theta} \overline{X}^a \right);
\label{fig3a}
\end{equation}

\noindent and\footnote{The evaluation of
Fig. 3(b). is somewhat involved; we owe the calculation to that of
Dorn and Otto in~\cite{dorn+otto}.}
\[
\exp(-{\rm Fig. ~3(b)}) = 
\]
\begin{equation}
(\alpha'g)^2\int_0^{2\pi} d\theta ~\left(\frac{\pi^2}{4}[N(0)]^2+\ln(\varepsilon)\right)~{\rm Tr} ~
{\cal P} \left( U[A]~ [D_a F_{bc}, F^{bc}] \partial_{\theta}
\overline{X}^a  \right),
\label{fig3b}
\end{equation}

\noindent all others being identically zero.  Due to the symmetry properties of the
region around $\theta \rightarrow \theta'$, this latter would be
zero in the abelian case but is non--zero for the case in
hand.

Adding (\ref{fig2}), (\ref{fig3a}) and
(\ref{fig3b}) we get the total 2-loop counterterm
\[
\exp(-\bigtriangleup S_2) =
\]
\[
- ig (\pi\alpha')^2 N_{div}^2
\int_0^{2\pi} d\theta~ {\rm Tr} ~ {\cal P} \left( U[A]~
\left[ -\frac{1}{2} D^2 D^bF_{ba} + \frac{4}{3} D^cD_aD^bF_{bc} -ig
[D^b F_{bc}, F^c_{~a}] \right] \partial_{\theta} \overline{X}^a \right)
\]
\[
-ig (\pi\alpha')^2 N_{div} \int_0^{2\pi} d\theta~N_{fin}~{\rm
Tr} ~ {\cal P}  \left( U[A]~ \left[ \frac{5}{3}D^cD_aD^bF_{bc}
-ig[D^b F_{bc}, F^c_{~a}] \right] \partial_{\theta} \overline{X}^a
\right)
\]
\begin{equation}
-(\alpha'g)^2 \ln(\varepsilon) \int_0^{2\pi} d\theta~ {\rm Tr}~{\cal
P} \left( U[A] ~[D_a F_{bc}, F^{bc}] \partial_{\theta} \overline{X}^a \right),
\label{counterterm2}
\end{equation}

\noindent where we have used the relation $D^2 F_{ab} = D_a D^cF_{cb}
- D_b D^c F_{ca} + 2ig[F_{ac}, F^c_{~~b}]$.  There are two points
about (\ref{counterterm2}) which must be considered: firstly, the $N_{div}^2$
term does not contribute to the $\beta$--function at all, else it
would itself be divergent; and, secondly, since (\ref{beta1}) gives the
equation of motion $D^a F^i_{ab} = {\cal O}(\alpha')$, the $N_{div}N_{fin}$ term is
effectively of ${\cal O}(\alpha'^3)$ and so can be dropped.
\newline
\newline
\[
\vbox{
         \beginlabels\refpos 75 605 {}
                     \put 105 595 {(a)}
                     \put 330 595 {(b)}
                     \put 105 470 {(c)}
                     \put 330 470 {(d)}
                     \put 170 500 {DDDF\partial_{\tau} \overline{X}}
                     \put 350 500 {F}
                     \put 482 500 {F}
                     \put 110 370 {DF\partial_{\tau} \overline{X}}
                     \put 260 370 {F}
                     \put 405 370 {DDF}
         \endlabels
         \epsfxsize=.85\hsize
         \centerline{\epsfbox{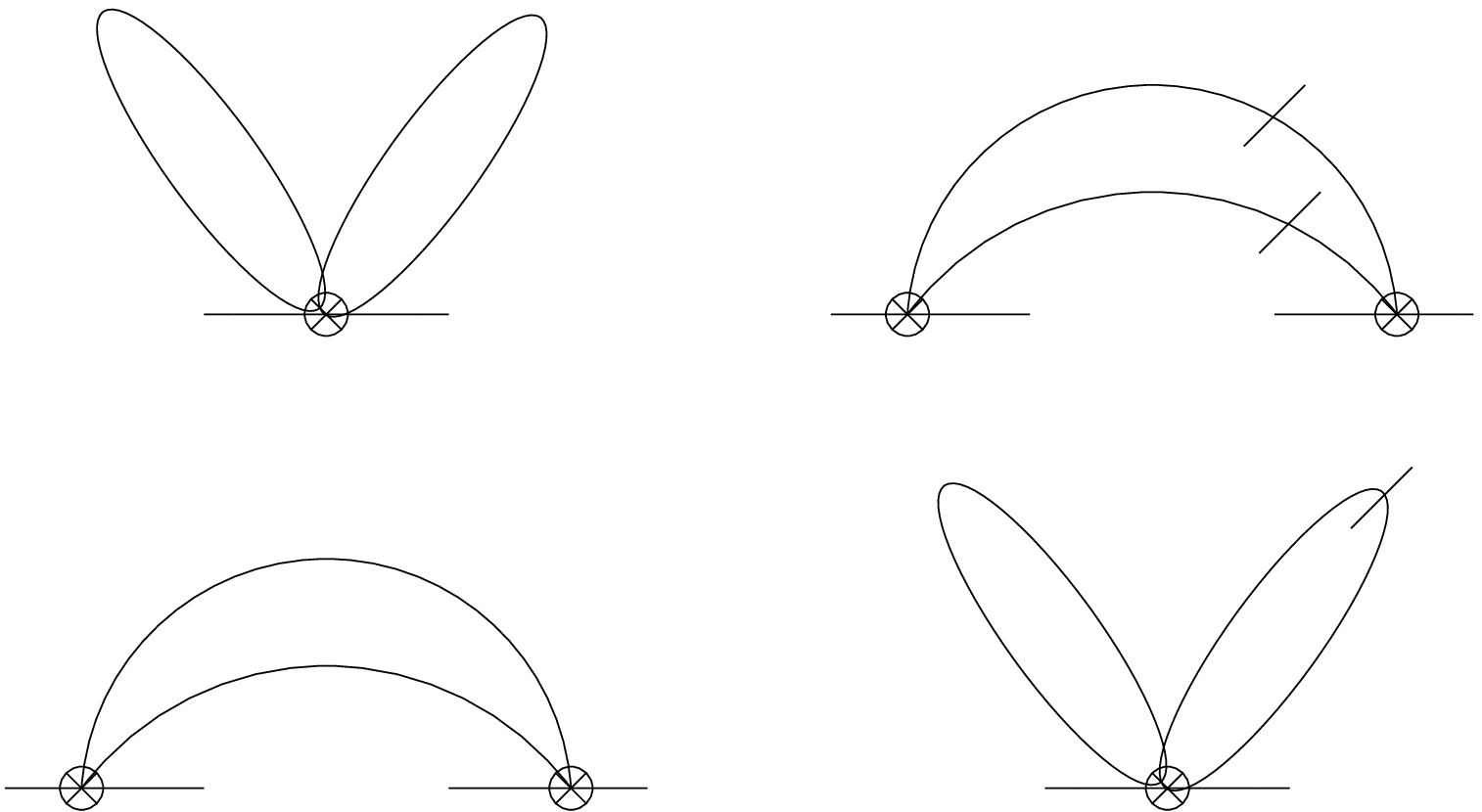}}
         \vspace{15pt}
         \centerline{\bf \footnotesize Figure 3.  Two--loop diagrams.}
         \vspace{-.1in}
         \centerline{\footnotesize Due to the
         symmetry properties involved, (c) and (d) are identically zero.}
            }
\]

Ignoring the first two terms in (\ref{counterterm2}), then, and
comparing with (\ref{counterexp}), we obtain the same
2--loop counterterm as~\cite{dorn+otto}:
\begin{equation}
\delta A_a =  (\alpha'g D^b F_{ba} + i (\alpha'g)^2
[D_a F_{bc}, F^{bc}] ) \ln(\varepsilon) + {\cal O}({\alpha'}^3).
\label{deltaA2}
\end{equation}

\noindent For a semisimple, compact group with $[t^i, t^j] =
ic^{ijk}t^k$, we have
\begin{eqnarray}
\beta_a^i &=& \alpha'gD^b F^i_{ba} - (\alpha'g)^2 c^{ijk} (D_a
F_{bc}^j) F^{kbc} + {\cal O}({\alpha'}^3) \nonumber\\
 &=&
\alpha'gD^b F^i_{ba} - 2(\alpha'g)^2 c^{ijk} D^b
(F_{bc}^j F^{kc}_{~~~a})+ {\cal O}({\alpha'}^3),
\label{beta2}
\end{eqnarray}

\noindent where we have used the Bianchi identity and have again dropped terms like $D^a F_{ab}^i \sim
{\cal O}(\alpha')$.  By setting $\beta^i_a = 0$, we obtain the
equation of motion for the Yang--Mills background gauge field, viz.
\begin{equation}
D^bF^i_{ba} - 2\alpha'g c^{ijk}D^b(F^j_{bc}F^{kc}_{~~~a})=0,
\label{eqnbos}
\end{equation}

\noindent the lowest order stringy correction to the Yang--Mills field
equation, at least within the bosonic theory.  We will see that for the
open superstring, the ${\cal O}({\alpha'}^2)$ term in the
$\beta$--function vanishes, the fermionic contribution cancelling the
two--loop bosonic divergence.

\subsection{The Open Superstring}

We have found the first order correction to the Yang--Mills equation
for the spacetime gauge field within open bosonic string theory.
Since this is precisely how the results which led to the BI action
were found, we should expect to be able to infer a non--abelian
generalisation of the BI (and the DBI) action from these calculations.  However, if
we are to apply our results to \emph{D-branes} we really should consider
the $\beta$--functions for the open \emph{super}string, since the
D-brane is embedded in the supersymmetric theory.  We have mentioned
above that, in the abelian case, the BI action is valid for the
superstring, as well as in the bosonic case~\cite{tseytlin:bi}.  In the
non--abelian case at hand, however, it is not at all obvious that the
fermionic degrees of freedom do not contribute to the
$\beta$--function.

Following~\cite{tseytlin:bi}, then, we define the
fermionic propagator, $K^{ab}(z,z')$, in a manner analogous to the
bosonic one: on the boundary, $K^{ab}(e^{i\theta}, e^{i\theta'}) =
-2\pi\alpha'\delta^{ab}K(e^{i\theta}, e^{i\theta'})$ where we take  
\begin{equation}
K(e^{i\theta}, e^{i\theta'})=-\frac{1}{\pi}\sum_{r=1/2}^{\infty} \sin
r\beta ~~e^{-\varepsilon r},
\end{equation}

\noindent the sum being over the half--integers since the fermionic
variable $\psi^a$ should be antiperiodic on the boundary of the disc.
Since we are dropping all ${\cal O}(\xi)$ terms in the expansion
(\ref{expansion}), there are only three extra fermionic diagrams to consider: those shown
in Fig. 4.  For the same reason that Fig. 3(b) is identically zero, so
is the one--loop diagram Fig. 4(a): $K(0) \sim \sin (0) = 0$.
There is thus no further contribution at the one--loop level.  It is
for this simple reason that the results of~\cite{callan+etal,leigh},
for the abelian case, carry
through to the supersymmetric theory.  The analysis of these papers is
at the one--loop level only, although is exact at this level for a
slowly--varying field strength; within this
approximation, then, the fermionic degrees of
freedom simply do not contribute, there being no non--zero one--loop
fermionic diagrams.

Since we are working to the two--loop order, our results will be valid
not just to lowest order in derivatives of the field strength
(although now, of course, these results will not be exact in
$\alpha'$).  The two--loop fermionic diagrams of Fig. 4 must, then, be
taken into account.  Fig. 4(b) is once again zero, so we are left with
Fig. 4(c) alone.  Now, it is easy to see that for the region of
interest, $\theta \rightarrow \theta'$, we have that $K^2 \sim
(\partial_{\theta} N)^2$, as should be expected from supersymmetry
considerations, and it is for this reason that
\begin{equation}
\exp(-{\rm Fig. ~4(c)}) = -(\alpha'g)^2\ln(\varepsilon)\int_0^{2\pi}
d\theta ~{\rm Tr} ~ {\cal P} \left( U[A]~ [D_a F_{bc}, F^{bc}]
\partial_{\theta} \overline{X}^a  \right),
\label{fermion}
\end{equation}

\noindent which exactly cancels the logarithmic divergence in (\ref{fig3b}).
After adding (\ref{fermion}) to the two--loop counterterm
(\ref{counterterm2}) and dropping the $N_{div}^2$ and
$N_{div}N_{fin}$ terms as before we find, then, that
\[
\exp(-\bigtriangleup S_2)_{{\rm total}} = 0 ~~\Rightarrow \]
\[\delta A_a^i = \alpha'g D^bF^i_{ba} \ln(\varepsilon) + {\cal O}({\alpha'}^3)
~~\Rightarrow 
\]
\begin{equation}
\beta_a^i = \alpha'gD^b F^i_{ba} + {\cal O}({\alpha'}^3).
\label{finalbeta}
\end{equation}

\noindent That is, the Yang--Mills equation is exact to ${\cal O}(\alpha')$, a
fact that has far--reaching ramifications for the non--abelian
generalisation of the BI action.\newline
\newline
\[
\vbox{
         \beginlabels\refpos 70 740 {}
                     \put 220 730 {(a)}
                     \put 75 570 {(b)}
                     \put 320 570 {(c)}
                     \put 270 597 {F}
                     \put 157 465 {DDF}
                     \put 332 465 {F}
                     \put 475 465 {F}
         \endlabels
         \epsfxsize=.85\hsize
         \centerline{\epsfbox{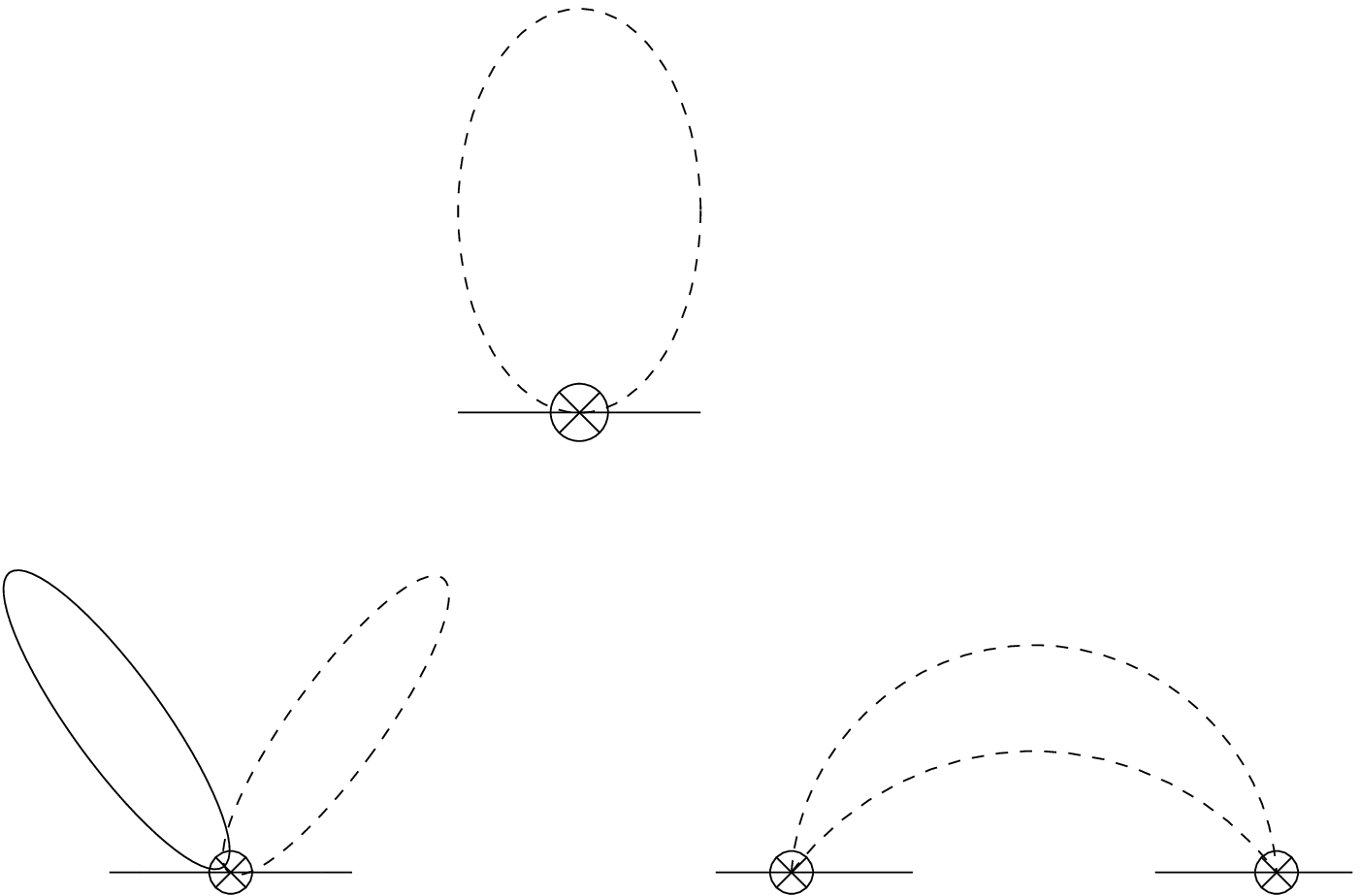}}
         \vspace{15pt}
         \centerline{\bf \footnotesize Figure 4.  One-- and two--loop
         diagrams fermionic contributions.}
         \vspace{-.1in}
         \centerline{\footnotesize The dashed line represents the
         fermionic propagator.}
         \vspace{-.1in}
         \centerline{\footnotesize Due to the
         symmetry properties involved, (a) and (b) are identically zero.}
            }
\]
 
It should be noted that during the preparation of this paper, it has
come to our attention that this result viz., the cancellation of the
${\cal O}(\alpha'^2)$ term, has been verified previously using
superspace techniques in~\cite{behrndt,behrndt:phd}.  Indeed,
three--loop diagrams are considered in this work, the implications of
which will be discussed below.

\section{The Non-Abelian Born-Infeld Action}

Our aim here is to find some non--abelian generalisation of the BI action
(\ref{lbi}), the equation of motion of which will have the same
solutions as the equation $\beta_b^i = 0$, with $\beta_b^i$ given in (\ref{finalbeta}).  We have shown this
latter to be just the usual Yang--Mills field equation, so it would
seem that we do not have much to go on.  This is not true, however:
the obvious first step is to simply replace the field strength in
(\ref{lbi}) with its non--abelian generalisation $F_{ab}^i t^i$, and to
replace the metric  with $\delta_{ab}{\cal I}$, ${\cal I}$ the unit
matrix over group space; giving $\sqrt{\det(\delta_{ab}{\cal
I}+2\pi\alpha' g F_{ab}^i t^i)}$.  Then, since the lagrangian must be a scalar, both in
spacetime as well as in the group, we must include some trace
(or determinant) operation over the group indices.  There has been
some recent~\cite{tseytlin}, and some not so
recent~\cite{argyres+nappi}, work which suggests two different such
trace operations; we will show that our results for the
superstring support the former.

We can immediately rule out certain possible generalisations
of the BI action.  That is, consider diagonalising the
field strength, $F_{ab}^i t^i = {\rm diag}(F_{ab}^1, F_{ab}^2, \ldots,
F_{ab}^N)$, and then performing the relevant trace operation.  We
should thereby obtain the sum of $N$ (ten--dimensional) BI actions.  In the
dimensionally reduced T--dual theory, to be discussed below, this would
correspond to $N$ separated and parallel D$p$-branes.  Now it is easy
to see that, only if the group trace operation occurs \emph{outside}
of the square root, will this be the case.  On these very general
grounds we must, then, have something of the form
\begin{equation}
{\cal L}_{NBI} = (2\pi\alpha'g)^{-2}~{\rm Tr} ~\left[ {\cal I}- \sqrt{\det
(\delta_{ab}{\cal I} + 2\pi\alpha' gF_{ab})}\right].
\label{lnbi1}
\end{equation} 

We can thus immediately exclude lagrangians with different
group trace structures, such as $\sqrt{{\rm Tr}\det (\delta_{ab}{\cal
I} + 2\pi\alpha'gF_{ab})}$, which was indeed proposed by
Hagiwara~\cite{hagiwara}, although not in the context of D-brane
theory, or $\sqrt{{\rm Det}\det (\delta_{ab}{\cal
I} + 2\pi\alpha'gF_{ab})}$, ${\rm Det}$ a determinant over group
space.  Indeed, the analysis of~\cite{argyres+nappi} also excludes
such lagrangians\footnote{In~\cite{argyres+nappi}, the low--energy
effective action for an open bosonic string charged under a $U(1)$
subgroup of the full--blown $U(N)$ Yang--Mills background field is
considered.  The effective action is inferred via BRST invariance, and
specifically excludes such terms as $({\rm Tr}F^2)^2$, which would be
generated if the group trace were to appear \emph{inside} of the
square root.}.  Moreover, this latter work also excludes the
obvious trace operation in (\ref{lnbi1}).  In the light of
these results, then, the only possible action, modulo certain terms to be discussed
below, is that proposed by Tseytlin in~\cite{tseytlin}:
\begin{equation}
{\cal L}_{NBI} =  (2\pi\alpha'g)^{-2}{\rm STr} ~ \left[ {\cal I}- \sqrt{\det (\delta_{ab}{\cal I} + 2\pi\alpha'gF_{ab})}\right],
\label{lnbi2}
\end{equation}

\noindent where ${\rm STr}(M_1,M_2,\ldots ,M_n) = \frac{1}{n!} \sum_{\pi}
{\rm Tr} \left( M_{\pi(1)} M_{\pi(2)} \ldots M_{\pi(n)} \right)$ denotes a
symmetrised group trace.  Note that we can further
define an antisymmetrised
group trace in a similar fashion: ${\rm ATr}(M_1, M_2, \ldots
,M_n) = \frac{1}{n!} \sum_{\pi} (-1)^{\pi} {\rm Tr} \left(
M_{\pi(1)}M_{\pi(2)}\ldots M_{\pi(n)} \right)$.  For reasons to be
discussed below, there is some
ambiguity here as to which representation the group trace is to be taken
over (cf.~\cite{tseytlin,gibbons2}).  Since the gauge field which
couples to the ends of the open string in the boundary action
(\ref{sbound}) \emph{must} be in the fundamental representation, however, the
field strength in (\ref{lnbi2}) must take values in this
representation also.  We will therefore require the symmetrised group
trace to be taken over the fundamental representation.

A na\"{\i}ve variation of the above action, ignoring
the matrix--ordering subleties, will just give the
obvious generalisation of the equation of motion (\ref{deltalbi}) for
the abelian case (e.g.~\cite{behrndt:phd}):
\begin{equation}
\frac{\delta {\cal L}_{NBI}}{\delta A_a^i} = {\rm STr} \left[
\sqrt{\det (\delta_{ab}{\cal I} + 2\pi\alpha'gF_{ab})} ({\cal I} -
(2\pi\alpha'g)F^2)^{-1b}_a \beta_b t^i \right],
\label{deltalnbi}
\end{equation}

\noindent with $\beta_a = \alpha'({\cal I} - (2\pi\alpha'g F)^2)^{-1bc}
D_{(b}F_{c)a}$ the natural generalisation of the
$\beta$--function for the abelian gauge field.  Under the ${\rm
STr}$ operation, it is clear that the different factors in
(\ref{deltalnbi}) will become mixed; and it is not obvious that this
equation of motion is equivalent to the statement $\beta_a^i = 0$, as
it is in the abelian case.  To ${\cal O}(\alpha'^2)$, however, and,
indeed, to ${\cal O}(\alpha'^3)$~\cite{behrndt:phd} the two viewpoints
are in fact equivalent.

At any rate, to this order, if we expand the spacetime determinant term--by--term it will be
obvious that the equation of motion following from (\ref{lnbi2}) is
identical to the statement $\beta_a^i = 0$.  Using the identity $\det M = \exp( {\rm tr}( \ln
M))$, with `${\rm tr}$' a trace over spacetime indices, we generate a sum of powers of $F_{ab}$:
\[
(2\pi\alpha'g)^{-2}\left[ {\cal I} -\sqrt{ \det ( \delta_{ab}{\cal I}
+ 2 \pi\alpha'g F_{ab} ) } \right] = 
\]
\begin{equation}
-\frac{1}{2}\left[ \frac{1}{2}F^2 + \frac{1}{3} (2\pi \alpha'g) F^3 +
\frac{1}{16} (2\pi\alpha'g)^2 \left( (F^2)^2 - 4 F^4 \right) \right] +
{\cal O}(\alpha'^3),
\label{detexpansion}
\end{equation}

\noindent where $F^2 = F_{ab}
F^{ab}$, $F^3 = F_{ab} F^{bc} F_c^{~~a} = \frac{1}{2} [F_{ab}, F^{bc}]
F_c^{~~a}$ and $F^4 = F_{ab} F^{bc} F_{cd} F^{da}$. Now it is easy to
see that, since $F_{ab}$ is antisymmetric in its spacetime indices,
the symmetrised trace will pick out the even powers of $F_{ab}$ from
this expansion only; and
an antisymmetrised trace, the odd powers only.  The lagrangian
(\ref{lnbi2}) thus contains even powers of $F_{ab}$ alone.  Indeed, it
is derived in~\cite{tseytlin} by assuming all odd powers of $F_{ab}$ to be
negligible.  Since $[D, D]F \sim
[F, F]F \sim F^3$, Tseytlin is led to assume
the $F^3$ term (and all higher order odd invariants) in
(\ref{detexpansion}) to be negligible within, that is, the approximation that the field strength $F_{ab}$
is slowly varying. Moreover, this
`slowly varying limit' is just an `abelian limit', in which the
matrices $F_{ab}$ are assumed to commute.  It is this `abelian
approximation' which is imposed by symmetrising the group trace in
(\ref{lnbi1}) to give (\ref{lnbi2}).

The question is, then, whether odd powers (or derivatives) of $F_{ab}$
should contribute to the non--abelian BI action.  The analysis
of~\cite{argyres+nappi}, in which it was found to be necessary to
introduce $F^3$ terms into the effective action, suggests that they
should; this being achieved by replacing ${\rm STr}$ with $({\rm STr}
+ i{\rm ATr})$ in (\ref{lnbi2}).  These results are for the \emph{bosonic}
string, however.  Indeed, our above analysis of the bosonic case supports
precisely this conclusion: the $D^b(F_{bc}F^c_{~a})$ term in (\ref{eqnbos}) is
just what would follow from a variation of an $F^3$ term in the
non--abelian BI action.

In the supersymmetric case, however, this term cancels with fermionic
contributions so, since the $F^3$ term cannot appear in the
non--abelian BI action, we must have ${\rm STr}$ as the correct trace
operation.  A three--loop analysis will contribute at ${\cal
O}(\alpha'^3)$, giving terms which should follow from the $F^4$ terms
in (\ref{lnbi2}):
\[ {\cal L}_{NBI} = -\frac{1}{2}\left[ \frac{1}{2}{\rm Tr}(F^2) + \right.\]
\begin{equation}
 +\left.\frac{1}{48}(2\pi\alpha')^2
{\rm Tr}\left(
F_{ab}F_{cd}F^{ab}F^{cd} +
2 F_{ab}F^{ab}F_{cd}F^{cd} -4 F_{ab}F^{bc}F_{cd}F^{da} -8
F_{ab}F^{bc}F_{ad}F^d_c \right)\right].
\label{lexp}
\end{equation}

\noindent Up to an overall factor, the coefficients of the $F^2$ and $F^4$ terms match those
found via stringy calculations in~\cite{tseytlin:bi}, in which the
absence of $F^3$ terms for the effective action of the superstring is
further confirmed.  To ${\cal O}(\alpha')$, (\ref{lexp}) just implies
the Yang--Mills equation, identical to $\beta_b^i = 0$ in
(\ref{finalbeta}).  This we take as our justification for the
non--abelian Born--Infeld action (\ref{lnbi2}).  The bottom line is
that both ${\rm Tr}$ and ${\rm ATr}$, when applied to the obvious
generalisation of the BI action will give $F^3$ terms (and higher
order odd invariants).  The fact that, at least the $F^3$ term should
not appear, implies that we must take ${\rm STr}$.

As mentioned above, a three--loop analysis is undertaken
in~\cite{behrndt,behrndt:phd}, in which it is verified that the $F^3$
term should be absent.  Moreover, it is shown that the ${\cal
O}(\alpha'^3)$ terms in the $\beta$--function agree with the equation
of motion following from (\ref{lexp}).  To this order, then, the
action (\ref{lnbi2}) is correct, as indeed is the view taken that the
vanishing of the $\beta$--function is equivalent to stationarity of
this action.

Our analysis suggests also an extension to the region of validity of
(\ref{lnbi2}), over and above that assumed in~\cite{tseytlin}, in that
we have shown the terms of lowest order in derivatives of the field
strength to be absent from the non--abelian BI action; this latter
should be correct not just for a \emph{slowly}--varying field
strength.  Whether \emph{all} higher order odd invariants should
vanish, the action depending solely on non--derivative terms, is a
question we cannot answer in the approach taken here without, that is,
performing higher--loop calculations.

\subsection{T-duality, D-branes and the Mass Term}

To see the appearance of D--branes we must consider applying
T--duality to our effective spacetime theory (\ref{lnbi2}).  That is,
we again consider compactifying the theory on a circle of radius $R$
in the $X^9$ direction~\cite{pol:tasi}.  Then, since the field $A_9$
must be pure gauge, we have the Wilson line
$A_9=-i\Lambda^{-1}\partial_9\Lambda={\rm
diag}\{\theta_1, \theta_2, \ldots , \theta_N\}/(2\pi R)$ and, under $X^9
\rightarrow X^9 +2\pi R$, the embeddings pick up a phase ${\rm
diag}\{e^{-i\theta_1}, e^{-i\theta_2}, \ldots, e^{-i\theta_N}\}$.  In
the dual theory, we have $\tilde{X}^9(\pi) - \tilde{X}^9(0) = (2\pi n
+ \theta_j - \theta_i)/\tilde{R}$, for a string with endpoints
in the state $\left| ij\right>$; up to a normalisation, the endpoint
in the state $i$ is at the position $\tilde{X}^9 = \theta_i \tilde{R}
= 2\pi\alpha' A_{9, ii}$.  This is just a theory of $N$ separated,
and parallel, D8-branes on the dual circle, the diagonal components
of $A_9$ specifying the positions of the branes along this circle.

Following~\cite{witten}, it is to be
understood that the dual coordinate $\tilde{X}^9$, interpreted
as a full--blown group matrix, specifies the transverse position of
the generic bound state of the $N$ D8-branes, at least when they are
all sitting on top of each other.  The process by which
the spacetime coordinates of such a `brane' become matrix--valued is by no
means clear, however.  Perhaps, if we were to compactify the ten--dimensional $U(N)$
theory on a manifold with a natural non--abelian structure (rather
than the trivial circle as above), some light might be shed on this.
Just as compactification on a circle gives a theory of $N$ separated
and parallel D8-branes, so compactification on a manifold with, say,
an $SU(2)$ structure should give a theory of $N/2$ separated and
parallel bound states of two D8-branes.  These latter will have
matrix--valued coordinates, the emergence of which might become
clearer by this process.

At any rate, we will simply suppose some compact directions, in which the
coordinates are assumed to be
matrix--valued: $X^{\alpha} = 2\pi\alpha' A^{\alpha}, ~~\alpha=p+1,
\ldots, 9$, where we have dropped the tilde from the dual coordinate.
For $A, B = 0, 1, \ldots, p$ denoting worldvolume directions, we have
$F_{A\alpha} =
(2\pi\alpha')^{-1}(\partial_AX_{\alpha}-i[A_A,X_{\alpha}])=(2\pi\alpha')^{-1}D_A X_{\alpha}$ and
$F_{\alpha\beta} = -i (2\pi\alpha')^{-2}[X_{\alpha}, X_{\beta}]$, since all fields are taken to depend on the worldvolume coordinates
alone (and where the gauge coupling constant $g$ has now been set to
unity).  Then using the identity (\ref{identity}), e.g.~\cite{tseytlin}:  
\[
{\cal L}_{NBI}  =
(2\pi\alpha')^{-2}{\rm STr}\left[ {\cal I} - \sqrt{\det(\delta_{\alpha\beta}{\cal I}  - i
(2\pi\alpha')^{-1} [X_{\alpha},  X_{\beta}])} \right.
\] 
\begin{equation}
\left.\times \sqrt{\det ( \delta_{AB}{\cal I} + D_A X_{\alpha}
(\delta_{\alpha\beta} - i (2\pi\alpha')^{-1} [X_{\alpha}, X_{\beta}])^{-1} D_B
X_{\beta} + 2\pi\alpha' F_{AB} ) } \right].
\end{equation}

\noindent This is the lagrangian relevant to the description of bound states of
D$p$-branes.  To lowest order in $[X_{\alpha}, X_{\beta}]$, we have
\[
{\cal L}_{NBI}=(2\pi\alpha')^{-2}{\rm STr}\left[{\cal I} - \left(
\sqrt{\det(  \delta_{AB}{\cal I} + D_A X_{\alpha} D_B X^{\alpha} +
2\pi\alpha' F_{AB})}\right.\right.
\]
\begin{equation}
\left.\left.\hfill-\frac{1}{4} (2\pi\alpha')^{-2}([X_{\alpha},
X_{\beta}])^2 \right) \right].
\label{pbrane}
\end{equation}

\noindent As is
to be expected, in the low--amplitude limit, the action (\ref{pbrane}) is
just the reduction from ten to $(p+1)$ dimensions of the
ten--dimensional $U(N)$ Yang--Mills theory, the potential
$V=\frac{1}{4}(2\pi\alpha')^2([X_{\alpha}, X_{\beta}])^2$.  

Since $N$ parallel D$p$-branes will interact via massive string states,
it is an interesting question as to whether this mass term will appear
within the non--abelian theory (\ref{pbrane}).  To this end we assume
the potential $V$ to vanish, in which case we can simultaneously
diagonalise the transverse coordinates $X^{\alpha}$, the diagonal
entries specifying the position of each brane: $X^{\alpha} = {\rm
diag}(X^{\alpha}_1, X^{\alpha}_2, \ldots, X^{\alpha}_N)$.  And this
is where the ambiguity arises as to which representation the symmetrised
trace is to be taken over.  It is usual to take the scalars in the
dimensionally reduced Yang--Mills theory to lie in the adjoint
representation, but we have stated above that the field strength, from
which they appear, lies in the fundamental.  Now consider the $SU(2)$
theory, which by rights should describe a bound state of \emph{two} D$p$-branes,
the entries in the diagonalised $X_{\alpha}$'s corresponding to the
positions of the two parallel and separated branes.  If we were to
take $X_{\alpha}$ to lie in the adjoint, it would be a $3 \times 3$
matrix, which surely cannot correspond in any way to just two branes.
If we consider the fundamental representation, however, then our
interpretation is consistent: the diagonal elements of the $2 \times
2$ matrices $X_{\alpha}$ can correspond to the positions of two
parallel branes.  Our statement above, then, that the symmetrised trace must
be taken over the fundamental representation, would seem to be
consistent.

With the transverse coordinates diagonalised, we have $V=0$; and there
are two distinct cases to consider.  Firstly, we can further
diagonalise the worldvolume field strength, as above.  Then $ D_A
X^{\alpha} \rightarrow \partial_A X^{\alpha}$ and ${\rm STr}
\rightarrow {\rm Tr}$, since everything is diagonal.  The action
(\ref{pbrane}) then reduces to the direct sum of $N$ DBI actions, with
the usual interpretation of $N$ separate and parallel
D$p$-branes. There is no mass term here.

A more interesting case would be to keep the field strength truly
non--abelian.  Then, although we should still have
the $N$ separate and parallel D$p$-branes (the transverse embeddings are
still diagonal), we should also see the interactions between the
branes emerging.  We assume the branes to
be stationary, in time and space, so $D_A X_{\alpha} =
-i[A_A,X_{\alpha}]$ where the world--volume gauge potential now has
off-diagonal components.  The simplest case is that of two D$p$-branes
with one transverse coordinate excited: $X = {\rm diag}(X_1, X_2)$.
The relevant gauge group, ignoring the overall abelian factor, is
$SU(2)$.  Then
\[
D_A X = -i (\bigtriangleup L) A_A^i \left( \begin{array}{cc} 0 &
t_{12}^i \\ -t_{21}^i & 0 \end{array}\right) = -i
\frac{(\bigtriangleup L)}{\sqrt{2}} \left( \begin{array}{cc} 0 &
W^*_A \\ -W_A & 0 \end{array}\right),
\]

\noindent with $t^i =
\sigma^i/2$, ${\rm Tr}(t^it^j)=\delta^{ij}/2$.  $\bigtriangleup L = X_2 - X_1$ is the separation of
the branes (the length of the strings between them), and we have set
$W^*_A = (A_A^1 - i A_A^2)/\sqrt{2}$ and $W_A = (A_A^1 + i A_A^2)/\sqrt{2}$.
Expanding the determinant in (\ref{pbrane}) to lowest order gives
\[
{\cal L}_{NBI}=(2\pi\alpha')^{-2}{\rm STr}\left[{\cal I}_2 - \left(
{\cal I}_2 + D_A X D^A X + \frac{1}{2} (2\pi\alpha')^2 F^2 \right.\right.
\]
\begin{equation}
\left.\left. -\frac{1}{2}[D_AXD_BXF^{BA}+F^{AB}D_BXD_AX] \right)^{1/2}\right].
\end{equation}

\noindent Substituting for $D_AX D^AX = ((\bigtriangleup L)^2
\left| W \right|^2)/2 ~{\cal I}_2$, where $\left| W \right|^2 =
W^A W^*_A = (A_A^1 A^{1A} + A_A^2 A^{2A})/2$, and to lowest order, we
have
\begin{eqnarray}
{\cal L}_{NBI} &=& -\frac{1}{4}{\rm Tr}(F^2) - \frac{1}{2}\frac{(\bigtriangleup
L)^2}{(2\pi\alpha')^2} \left| W \right|^2 \nonumber\\
 &=&
-\frac{1}{4}{\rm Tr}(F^2) - \frac{1}{2}M^2 \left| W \right|^2,
\end{eqnarray}

\noindent since ${\rm STr} (t^i t^j) = {\rm Tr} (t^i t^j)$ and ${\rm
STr} ({\cal I}_2) = {\rm Tr} ({\cal I}_2) = 2$.  $M = (\bigtriangleup
L)/(2\pi\alpha')$ is the mass of the string stretching between
the two branes.  This analysis follows through with more than one
compact direction, in which case $(\bigtriangleup L)^2 =
(X^2_{\alpha}-X^1_{\alpha})(X^{2\alpha}-X^{1\alpha})$.  Note that the
cross term $[D_AX D_BX F^{BA}+F_{AB} D^BX D^AX]$ drops out under the
${\rm STr}$ operation.  The equations of motion are thus
\begin{equation}
\left.\begin{array}{ccl} D^AF_{AB}^3 & = & 0 \\ [.05in]
D^AF_{AB}^i & = & M^2A_B^i \end{array}\right\}
\label{eqnsmot}
\end{equation}

\noindent where now $i=1,2$ only.  We can see that the mass term has appeared in
the form of a massive Yang--Mills theory in the $\{1,2\}$ group
directions.  As should be expected, in this case gauge invariance is
not realised in any obvious way.

There is a nice check of this result:
following~\cite{dorn:review}, we can consider applying T--duality to
the $\beta$--function itself.  With $\beta_b = \alpha' D^aF_{ab}$ and
$X^9 \equiv X = 2\pi\alpha'A^9$, we have
\begin{equation}
\beta_B = \alpha' D^A F_{AB} + \frac{i}{4\pi^2\alpha'}[X, D_BX]
= 0, \label{betaB}
\end{equation}
\[
\beta_9 = \frac{1}{2\pi}D^AD_A X=0,\nonumber
\]

\noindent where, as above, we have assumed $X$ to be diagonalised.
For static branes, we have
\begin{equation}
\frac{1}{\sqrt{2}}(D^AF_{AB})_{ij} - M^2 \left( \begin{array}{cc} 0 &
W^*_B \\ W_B & 0 \end{array}\right) = 0.
\end{equation}

\noindent This implies $D^AF_{AB}^3 =0$ and $D^AF_{AB}^i = M^2A_B^i$
which are precisely the equations (\ref{eqnsmot}).  To summarise, the
application of T--duality to
the ten--dimensional $\beta$--function gives an equation of motion which, at least
to lowest order, is just that obtained from the T--dual of the
ten--dimensional non--abelian BI theory.

Moreover, we have shown how the non--abelian BI theory describes two
parallel and interacting D-branes: to lowest order, this description
is just that of a massive Yang--Mills field (in two of the group
directions).

\subsection{The (Anti)--Self--Dual Case}  

Unfortunately, an analysis of the non--abelian BI theory
(\ref{lnbi2}), or its dimensionally reduced version (\ref{pbrane})
in the generic case, and for anything but the lowest order terms, is a
difficult problem.  For multiple D4-branes
with an (anti--)self--dual field strength, however, a considerable
simplification occurs, due to the fact that the determinant can be
written as a complete square.  We consider the purely magnetic case,
$F_{0A} = 0, ~~A=1, \ldots, 4$.  Then~\cite{towns}
\begin{eqnarray}
\det (\delta_{AB} + 2\pi\alpha'F_{AB}) &=& 1 + \frac{1}{4}(2\pi\alpha')^2F^2 +\frac{1}{2}(2\pi\alpha')^2\tilde{F}^2+ \frac{1}{16}(2\pi\alpha')^4(F\cdot
\tilde{F})^2 \nonumber\\
 &=&
(1\mp \frac{1}{4}(2\pi\alpha')^2{\rm tr}F\tilde{F})^2-\frac{1}{4}(2\pi\alpha')^2{\rm tr}(F\mp \tilde{F})^2,
\end{eqnarray}

\noindent where $\tilde{F}_{AB}$ is the dual of $F_{AB}$.  Setting all
transverse coordinates to zero, and for an (anti--)self--dual field
strength, $F_{AB} = \pm \tilde{F}_{AB}$, we then have
\begin{equation}
{\cal L}_{NBI} = (2\pi\alpha')^{-2} {\rm STr} \left[ {\cal I} -
\sqrt{\left( {\cal I} + \frac{1}{4}(2\pi\alpha')^2F^2\right)^2}\right]
 =
-\frac{1}{4}{\rm Tr}(F^2),
\end{equation}

\noindent which is just the lagrangian for the linearised
(Yang--Mills) theory.  This has in fact already been noted
in~\cite{hash}.  The self--duality condition is, as usual,
solved by instanton configurations.  In this case, then, the
non--abelian BI action reduces to that of Yang--Mills theory.  As
explained in~\cite{towns}, the energy of solutions in this case is an
absolute minima: they are BPS states.
In fact, this will be the case for any situation in which the determinant can be
written as a complete square.  Most of the analysis of~\cite{towns}
will, then, carry through to the non--abelian case, at least for the
BPS states themselves, for which the energy bound is saturated.  It
would seem, then, that the BPS states of the non--abelian BI theory
are just those of its linearised version; it would be interesting to
analyse this statement in more detail.

\section{Discussion}

The $\beta$--function for the non--abelian gauge field in open bosonic
string theory suggests an extension of Tseytlin's recent proposal for
the non--abelian BI action to be necessary.  This is not true for the
superstring, however; and since it is this latter case which is
relevant to the study of bound states of D-branes, we effectively
verify Tseytlin's proposal.  We have shown how this proposal, and no
other (no obvious others at least), gives an
equation of motion compatible with the equation $\beta_b^i=0$ for the
superstring.  We have discussed the application of T--duality to this
action, giving the action relevant to bound states of D-branes; and
have shown how the interaction between two separated and parallel
branes appears in the theory.  Although we have an action, it is hard
to know quite what to do with it, in all but the simplest of cases.
That is, when the determinant can be written as a complete square.
For D4-branes, with an (anti--)self--dual field strength, the action
reduces to the usual Yang--Mills action, the instanton solutions of
which are then BPS states in the non--abelian BI theory.  A detailed
analysis of the action for D0-branes would be most welcome; this providing
information as to corrections to the usual (dimensionally reduced)
Yang--Mills action of M(atrix) theory.\newline\newline

\noindent {\bf Acknowledgments}

One of the authors (DB) would like to thank Philip Argyres and Harald
Dorn for electronic correspondence, Gary Gibbons and George
Papadopoulos for enlightening discussions, and Miguel Costa and Steve Hewson for
keeping me going.

\newpage

\appendix
\section{Appendix}

\noindent The expansion, (\ref{expansion}), was derived in the early
days of string theory when the role of the Wilson loop, providing a
bridge between Yang--Mills and string theory, was emphasised;
for a review see, e.g.,~\cite{dorn:wilsonloops}.  We present here a derivation of the bosonic sector of (\ref{expansion}) for completeness, following the
analysis of~\cite{gervais+neveu}; the fermionic pieces follow easily.  Consider, then, an arbitrary closed curve
$X^a (\tau), 0 \leq \tau \leq 1$, and let $W[X] = {\cal P} \left( \exp ig
\int^1_0 A_b(X) dX^b \right) = {\cal P} \left( U[A] \right)$.  Then
\[W[\overline{X}+\xi] - W[\overline{X}] = \int^1_0 d\tau \left. \frac{\delta W[X]}{\delta X^a(\tau)}
\right|_{X=\overline{X}} \xi^a(\tau) \]
\begin{equation}
+ \frac {1}{2!}\int^1_0 d\tau_1 \int_0^1 d\tau_2
\left. \frac{\delta^2 W[X]}{\delta X^a(\tau_1) \delta X^b(\tau_2)}
\right|_{X=\overline{X}} \xi^a(\tau_1) \xi^b(\tau_2) + {\cal O}(\xi^3)
\label{Wexp}
\end{equation}

\noindent To compute the functional derivatives, we expand the embedding
\begin{equation}
X^a(\tau) =\lim_{M \rightarrow \infty} X^a_{(M)} (\tau) =  \lim_{M \rightarrow \infty} \sum^M_{q=1} \alpha^a_q f_q
(\tau),
\end{equation}

\noindent where
\begin{equation}
\alpha^a_q = \int^1_0 d\tau f_q(\tau) X^a(\tau).
\end{equation}

\noindent $\{f_q(\tau)\}$ are a set of `equally dense' functions such that
\begin{eqnarray}
f_q(\tau + 1) = f_q(\tau) 
  &,&
~~\int^1_0 d\tau f_p(\tau) f_q(\tau) = \delta_{pq},\nonumber\\
\sum^M_{q=1} f_q(\tau_1) f_q(\tau_2)
  &=&
\delta_{(M)}(\tau_1 - \tau_2),\label{sumff}\\
\lim_{M \rightarrow \infty} \frac{1}{M}\delta_{(M)}(\tau_1 - \tau_2)
  &=&
\delta (\tau_1 - \tau_2),\nonumber
\end{eqnarray}

\noindent Working in D--dimensional spacetime, the functional $W[X]$ is considered as the limit of a function of $M \times D$ variables:
\begin{equation}
W[X] = \lim_{M \rightarrow \infty} W_{(M)}(\alpha^{a_1}_1 \ldots
\alpha^{a_M}_M) \equiv \lim_{M \rightarrow \infty} W[X_{(M)}],
\end{equation}

\noindent and the functional derivatives are defined in terms of
partial derivatives via
\begin{equation}
\frac{\delta}{\delta X^a(\tau)} = \lim_{M \rightarrow \infty}
\sum^M_{q=1} f_q(\tau) \frac{\partial}{\partial \alpha^a_q}.
\label{funcder}
\end{equation}

\noindent To compute the partial derivatives of $W[X]$, we must first
expand the path ordered exponential:
\[W[X] = ig\int_0^1 d\tau A_b(X)\partial_{\tau}X^b
+ \frac{(ig)^2}{2}\int_0^1 d\tau_1 \int_0^{\tau_1} d\tau_2
[A_b(X(\tau_1)),A_c(X(\tau_2))] \partial_{\tau_1}X^b
\partial_{\tau_2}X^c \]
\begin{equation}
+\frac{(ig)^3}{2} \int_0^1 d\tau_1 \int_0^{\tau_1} d\tau_2 \int_0^{\tau_2}
d\tau_3 A_b(X(\tau_1))[A_c(X(\tau_2)),A_d(X(\tau_3))] \partial_{\tau_1}X^b
\partial_{\tau_2}X^c \partial_{\tau_3}X^d +{\cal O}(e^4),
\label{pathorder}
\end{equation}

\noindent needing the terms through ${\cal O}(g^2)$ to compute the first
derivatives of $W[X]$, and the terms through ${\cal O}(g^3)$for the
second derivatives.  We need compute the first partial derivative
only, however, this being
\begin{equation}
\frac{\partial W[X_{(M)}]}{\partial \alpha^a_q} = {\cal P} \left(
U[A] ~ig \int_0^1 d\tau f_q  F_{ab}(X_{(M)}) \partial_{\tau}X^b_{(M)} \right).
\label{partial1}
\end{equation}

\noindent Using the definition (\ref{funcder}), the property
(\ref{sumff}), and taking the limit $M \rightarrow \infty$ we then
find
\begin{equation}
\frac{\delta W[X]}{\delta X^a (\tau)} = {\cal P} \left( U[A] ~ig
F_{ab}(X(\tau)) \partial_{\tau}X^b \right).
\label{func1}
\end{equation}

\noindent The second functional derivative can be found easily from
(\ref{func1}), by the repeated use of (\ref{funcder}):
\[\frac{\delta^2 W[X]}{\delta X^b (\tau_2) \delta X^a (\tau_1)} =
\lim_{M \rightarrow \infty} \sum^M_{q=1} f_q(\tau_2)
\frac{\partial}{\partial \alpha^b_q} \left( \frac{\delta W[X]}{\delta
X^a (\tau_1)} \right) = \]
\[ig{\cal P} \left( U[A] \left[ D^b F_{ac} (X(\tau_1))
\partial_{\tau_1} X^c \delta (\tau_1 - \tau_2) + F_{ab}
\partial_{\tau_1} \delta (\tau_1 - \tau_2) \right. \right.\]
\begin{equation}
\left. \left. ie F_{ac}(X(\tau_1)) \partial_{\tau_1} X^c F_{bd}(X(\tau_2))
\partial_{\tau_2} X^d \right] \right),
\label{func2}
\end{equation}

\noindent where the endpoint contributions (at $\tau = 0,1$), which
vanish anyway for a closed path, have been ignored.  We note that the
full non--abelian field strength and its covariant
derivative appear in (\ref{func1}) and (\ref{func2}),
respectively, due to the path ordering operation which introduces the
relevant commutator terms, these being specific to the non--abelian
case.  Working to ${\cal O}(g)$, we now evaluate (\ref{func1}) and
(\ref{func2}) at $X = \overline{X}$ and substitute into (\ref{Wexp}), giving
\[W[\overline{X}+\xi] - W[\overline{X}] = \]
\begin{equation}
ig \int_0^1 d\tau {\cal P} \left( U[A]  \left[ F_{ab} (\overline{X})
\xi^a \partial_{\tau} \overline{X}^b  + \frac{1}{2!} D_b F_{ac}
(\overline{X}) \xi^a \xi^b \partial_{\tau} \overline{X}^c +
\frac{1}{2!} F_{ab} (\overline{X}) \xi^a \partial_{\tau}
\xi^b \right] \right).
\end{equation}

\noindent To ${\cal O}(\xi^2)$, then, this is just the expansion
(\ref{expansion}), as promised.  The higher order terms are generated by a
straightforward generalisation of this result noting that, at each
order, we will have two terms: one of the form $D \ldots DF \xi \ldots
\xi \partial_{\tau} X$; and one of the form $F \xi \ldots \xi
\partial_{\tau} \xi$.

\newpage
\bibliography{nonab_BI}

\begin{thebibliography}{10}

\bibitem{tseytlin}
A.~A. Tseytlin.
\newblock {\em Nucl. Phys.}, B501:41, 1997.
\newblock hep-th/9701125.

\bibitem{dai+leigh+pol}
J.~Dai, R.~G. Leigh, and J.~Polchinski.
\newblock {\em Mod. Phys. Lett.}, A4:2073, 1989.

\bibitem{pol:dbrane}
J.~Polchinski.
\newblock {\em Phys. Rev. Lett.}, 75:4724, 1995.
\newblock hep-th/9510017.

\bibitem{pol:tasi}
J.~Polchinski.
\newblock {TASI} {L}ectures on {D}-{B}ranes.
\newblock hep-th/9611050.

\bibitem{born+infeld}
M.~Born and L.~Infeld.
\newblock {\em Proc. Roy. Soc.}, A144:425, 1934.

\bibitem{callan+etal}
A.~Abouelsaood, C.~G. Callan, C.~R. Nappi, and S.~A. Yost.
\newblock {\em Nucl. Phys.}, B280:599, 1987.

\bibitem{leigh}
R.~G. Leigh.
\newblock {\em Mod. Phys. Lett.}, A4:2767, 1989.

\bibitem{dorn:dbranes}
H.~Dorn.
\newblock {\em Nucl. Phys.}, B494:105, 1997.
\newblock hep-th/9612120.

\bibitem{argyres+nappi}
P.~C. Argyres and C.~R. Nappi.
\newblock {\em Nucl. Phys.}, B330:151, 1990.

\bibitem{dorn+otto}
H.~Dorn and H.-J. Otto.
\newblock {\em Z. Phys.}, C32:599, 1986.

\bibitem{witten}
E.~Witten.
\newblock {\em Nucl. Phys.}, B460:335, 1996.
\newblock hep-th/9510135.

\bibitem{banks+etal}
T.~Banks, W.Fischler, S.~H. Shenker, and L.~Susskind.
\newblock {\em Phys. Rev.}, D55:5112, 1997.
\newblock hep-th/9610043.

\bibitem{callan+etal2}
C.~G. Callan, D.~Friedan, E.~J. Martinec, and M.~J. Perry.
\newblock {\em Nucl. Phys.}, B262:593, 1985.

\bibitem{callan+thorlacius}
C.~G. Callan and L.~Thorlacius.
\newblock Sigma {M}odels and {S}tring {T}heory.
\newblock In A.~Jevicki and C.-I. Tan, editors, {\em Particles, Strings and
  Supernovae: Proc. TASI 1988}, page 795. World Scientific, 1989.

\bibitem{alv-gau+freedman}
L.~Alvarez-Gaum{\'{e}} and D.~Z. Freedman.
\newblock {\em Ann. Phys.}, 134:85, 1981.

\bibitem{tseytlin:bi}
A.~A. Tseytlin.
\newblock {\em Nucl. Phys.}, B276, B291(E):391, 876(E), 1986, 1987(E).

\bibitem{gibbons}
G.~W. Gibbons.
\newblock {\em Nucl. Phys.}, B514:603, 1998.
\newblock hep-th/9709027.

\bibitem{gervais+neveu}
J.~L. Gervais and A.~Neveu.
\newblock {\em Nucl. Phys.}, B153:445, 1979.

\bibitem{dorn:wilsonloops}
H.~Dorn.
\newblock {\em Fortschr. Phys.}, 34:11, 1986.

\bibitem{frad+tseytlin}
E.~S. Fradkin and A.~A. Tseytlin.
\newblock {\em Phys. Lett.}, 163B:123, 1985.

\bibitem{behrndt}
K.~Behrndt.
\newblock Open {S}uperstring in {N}on--{A}belian {G}auge {F}ield.
\newblock In H.~J. Kaiser, editor, {\em Proc. XXIII Int. Symp. Ahrenshoop
  1989}, page 174. Akademie der Wissenschaften der DDR, 1989.

\bibitem{behrndt:phd}
K.~Behrndt.
\newblock {\em Untersuchung der {W}eyl--{I}nvarianz im {V}erallgemeinerten
  $\sigma$--{M}odell f{\"{u}}r {O}ffene {S}trings}.
\newblock PhD thesis, Humboldt--Universit{\"{a}}t zu Berlin, 1990.

\bibitem{hagiwara}
T.~Hagiwara.
\newblock {\em J. Phys.}, A14:3059, 1981.

\bibitem{gibbons2}
G.~W. Gibbons.
\newblock Wormholes on the {W}orld {V}olume.
\newblock hep-th/9801106.

\bibitem{dorn:review}
H.~Dorn.
\newblock Non--{A}belian {G}auge {F}ield {D}ynamics on {M}atrix {D}-{b}ranes in
  {C}urved {S}pace and {T}wo--{D}imensional $\sigma$-{M}odels.
\newblock hep-th/9712057.

\bibitem{towns}
J.~P. Gauntlett, J.~Gomis, and P.~K. Townsend.
\newblock {\em J. High Energy Phys.}, 01:003, 1998.
\newblock hep-th/9711205.

\bibitem{hash}
A.~Hashimoto.
\newblock {\em Phys. Rev.}, D57:6441, 1998.
\newblock hep-th/9711097.

\end{thebibliography}

\end{document}